\documentclass{PoS}

\usepackage{graphics}
\usepackage{fancyvrb}
\usepackage{xcolor}
\usepackage{xspace}
\usepackage{tikz} 
\usepackage{amsmath}
\usetikzlibrary{chains,shapes,fit,calc}
\usetikzlibrary{positioning,arrows}

\usepackage[utf8]{inputenc}
\usepackage{bbding,wasysym}
\usepackage{array}
\usepackage{listings}

\definecolor{unigrey}{rgb}{0.42,0.43,0.44}
\definecolor{uniblue}{rgb}{0,.59,.51}
\usepackage{multirow,rotating}
\definecolor{lightBlue}{rgb}{0.0,0.4,0.6}

\usepackage{bbding}

\newcommand{\DR}{\ensuremath{\overline{\rm{DR}}}}

\newcommand{\SARAH}{{\tt SARAH}\xspace}
\newcommand{\SPheno}{{\tt SPheno}\xspace}

\newcommand{\yes}{\Checkmark}
\newcommand{\no}{\XSolidBrush}

\title{Review of Calculators for BSM Higgs bosons}

\ShortTitle{Review of Calculators for BSM Higgs bosons}

\author{\speaker{Florian Staub}\thanks{A footnote may follow.}\\
        Institute for Theoretical Physics (ITP), Karlsruhe Institute of Technology, Engesserstra{\ss}e 7, D-76128 Karlsruhe, Germany\\
Institute for Nuclear Physics (IKP), Karlsruhe Institute of Technology, Hermann-von-Helmholtz-Platz 1, D-76344 Eggenstein-Leopoldshafen, Germany\\
        E-mail: \email{florian.staub@kit.edu}}

\abstract{We have reached a new era of particle physics in which the properties of the Higgs boson, in particular its mass, turned into precision observables.
Therefore, it is necessary to have accurate predictions of these properties in models for new physics. I give an overview of available tools for supersymmetric 
and non-supersymmetric models which are publically available. Afterwards, the main idea behind generic tools to study a large variety of models is summarised. 
Also some remarks about the validity of checks for perturbative unitarity, which are often applied for non-supersymmetric models, are given.}

\FullConference{
Prospects for Charged Higgs Discovery at Colliders - CHARGED2018\\
		25-28 September 2018\\
		Uppsala, Sweden}

\begin{document}

\section{Introduction}
The Large Hadron Collider has opened a new era of precision physics. Following the discovery of the Higgs, the measurement of its properties -- in particular its mass -- have now been performed with an astonishing precision. This is interesting because a precise determination of the Higgs properties is of crucial importance in understanding the fate of the Standard Model  and is especially sensitive to new physics beyond the SM (BSM). Therefore, public tools are developed to get precise prediction for the properties of the SM-like Higgs but also for other fundamental scalars in BSM models. \\
Before we discuss the different tools, we want to consider the question what are suitable input parameters for a BSM model and what the output of the tools should be. We can distinguish two cases:
\begin{enumerate}
 \item {\bf Supersymmetric Theory}: \\
 In a SUSY theory, the Lagrangian parameters are the input. Therefore, it is necessary to calculate the masses of the scalars from these parameter including important higher order corrections. The calculation of other properties like decay widths, branching ratios or production rates are only a second step.
 \item {\bf Non-Supersymmetric Theory}: \\
 In Non-SUSY models it is common to use physical masses/angles as input. Therefore, these tools can focus immediately on the calculation of decays and other features. 
\end{enumerate}
The reason for this big difference between SUSY and non-SUSY models are the strong constraints on the shape of the scalar potential in SUSY models. While in non-SUSY models the scalar self-interactions are free parameters, this is not the case in SUSY theories where those terms are fixed by $F$- and $D$-term conditions. Thus, a non-SUSY model has usually a sufficient number of counter-terms to perform a full on-shell (OS) calculation, while this is not the case for a SUSY model.

\section{SUSY Tools}
The tasks of a 'Spectrum generator' are 
\begin{enumerate}
 \item Perform a RGE running if an unified/GUT scenario is assumed
 \item Calculate the mass spectrum including important radiative corrections
 \item Optionally: calculate decays, flavour observables or other information for a given parameter point
\end{enumerate}
Note, I focus here on tools calculating Higgs (and SUSY) masses. For other purposes tools like {\tt SDECAY}/{\tt SUSY-HIT}, {\tt SFOLD}, {\tt Higlu}, {\tt SuSHi}, \dots exist.

\subsection{MSSM}
\begin{figure}[tb]
 \begin{center}
  \includegraphics[width=0.66\linewidth]{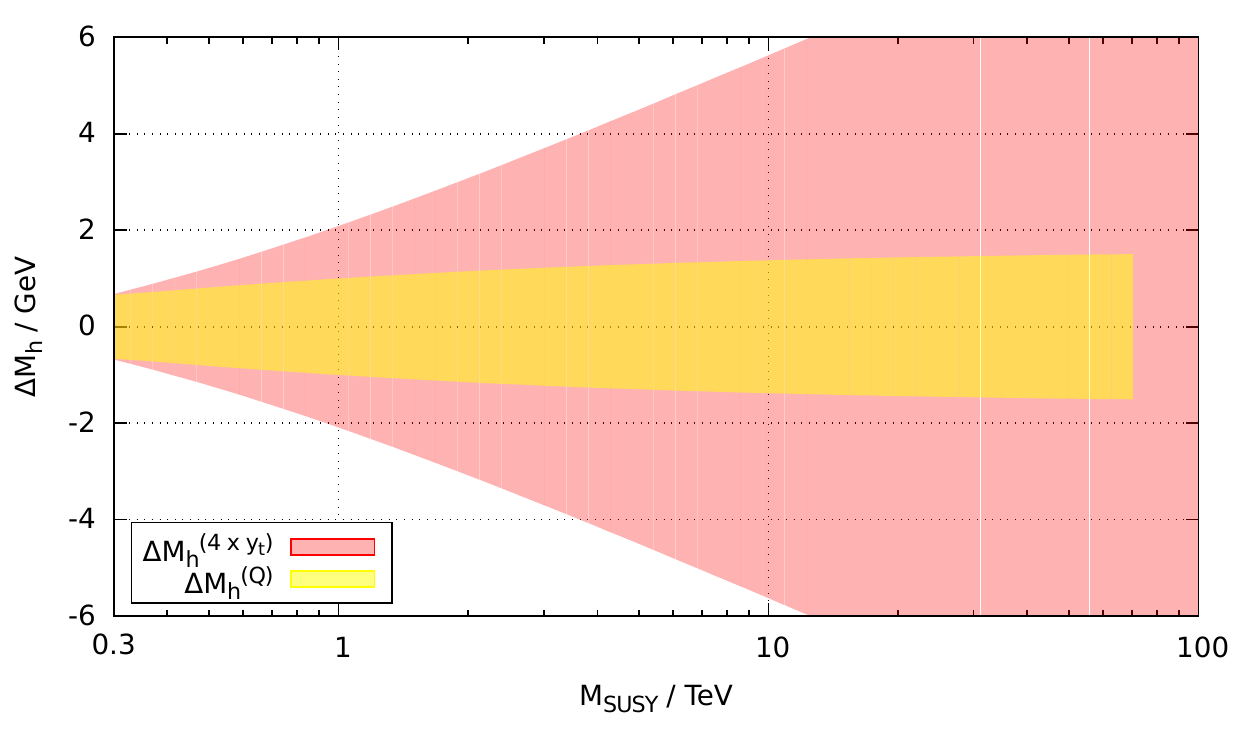}
 \end{center}
\caption{Theoretical uncertainty of the light Higgs mass for a fixed order calculation as function of the SUSY scale. Plot taken from \cite{Athron:2016fuq}. } 
\label{fig:uncertainty}
\end{figure}

Well established codes available for the MSSM are:
\begin{itemize}
 \item {\bf (Standard) Spectrum Generators}: {\tt SoftSUSY} \cite{Allanach:2001kg}, {\tt SPheno}\cite{Porod:2011nf,Porod:2003um}, {\tt Suspect}\cite{Djouadi:2002ze}
 \item {\bf Higgs focused tools}: {\tt FeynHiggs}\cite{Heinemeyer:1998yj}, {\tt CPsuperH}\cite{Lee:2003nta}, {\tt H3M}/{\tt Himalaya}\cite{Harlander:2017kuc} \\[5mm]
\end{itemize} 
However, the LHC results had some impact on this landscape:
\begin{itemize}
\item These tools are well established since many years but need(ed) to be revised because of LHC results. 
\item New tools appeared: {\tt SusyHD}\cite{Vega:2015fna}, {\tt MhEFT}\cite{Lee:2015uza} 
\end{itemize}
The reason for this is that most MSSM spectrum generators use(d) the same approach to get the SUSY and Higgs masses:
\begin{enumerate}
 \item The matching to the SM model was performed as in Ref.~\cite{Pierce:1996zz}, i.e.
 \begin{enumerate}
  \item The DR values of the gauge and Yukawa couplings $g_i$, $Y_i$ at the scale $Q=M_Z$ are derived from $G_F, \alpha_{ew}, \alpha_S, M_Z, m_q, m_l$ at the one-loop level
  \item SUSY RGEs are used between $M_Z$ and $M_{SUSY}$
 \end{enumerate}
 Of course, this is only a good approximation as long as $M_{\rm SUSY} \simeq M_Z$ holds
 \item The SM-like Higgs mass is calculated at the scale $Q=M_{SUSY}$ in full MSSM usually at two-loop level. This leads to a significant rise 
 in the theoretical uncertainty with increasing $M_{\rm SUSY}$ as shown in Fig.~\ref{fig:uncertainty}.
\end{enumerate}
This has caused significant improvements in the codes: the SUSY particles are now usually decoupled at the SUSY scale with SM RGEs between $M_Z$ and $M_{\rm SUSY}$. Moreover, the 
Higgs mass is not longer calculated at the fixed order but EFT techniques are used:
 \begin{enumerate}
  \item Match the SM and MSSM at $M_{\rm SUSY}$ to obtain $\lambda_{\rm SM}(M_{\rm SUSY})$
  \item Run $\lambda_{\rm SM}$ to $m_t$
  \item Calculate $m_h$ at $m_t$ with SM corrections
 \end{enumerate}  
The common models considered as EFT are usually the SM or 2HDMs. An overview about the Tools for the MSSM and their features is given in Tab.~\ref{tab:MSSM}. 

\begin{table}[tb]
\begin{center}
\begin{tabular}{|p{2.0cm}| >{\centering\arraybackslash}p{1.75cm} >{\centering\arraybackslash}p{1.75cm}  >{\centering\arraybackslash}p{1.75cm} >{\centering\arraybackslash}p{1.75cm} >{\centering\arraybackslash}p{1.75cm} >{\centering\arraybackslash}p{1.75cm}>{\centering\arraybackslash}p{1.75cm} |}
\hline 
                    & {\tt CPsuperH}   &  {\tt FeynHiggs}    &   {\tt MhEFT}     & {\tt SoftSUSY}           & {\tt SPheno}      & {\tt Suspect}                &  {\tt SusyHD}                     \\
\hline  
\hline 
\multicolumn{8}{|c|}{General} \\
\hline
GUT model           &  \no             &  \no                &   \no             & \yes                      &    \yes          &  \yes                         & \no \\
CPV                 &  \yes            &   \yes              &   \no             & partially                 & partially        & \no                           & \no \\
\hline         
\hline
\multicolumn{8}{|c|}{Heavy SUSY} \\
\hline
supported           & \yes             & \yes                & only             & -               & \yes             & \no                           & only\\
EFT                 & SM               & SM, THDM            & THDM             & -                         & SM               & -                             & SM  \\
\hline
\hline 
\multicolumn{8}{|c|}{Other observables} \\
\hline 
Decays              & \yes             & \yes               & \yes              & \yes                      & \yes              & \yes                          & \no \\
Flavour             & \no              & \yes               & \no               & \no                       & \yes              & \yes                          & \no \\
$g-2$               & \yes             & \yes               & \no               & \no                       & \yes              & \yes                          & \no \\
Oblique P.          & \no              & \yes     & \no               & \no                       & \yes              & \no                           & \no \\
EDMs                & \yes             & \yes               & \no               & \no                       & \yes              & \yes                          & \no \\
\hline
\end{tabular}
\end{center}
\caption{Public tools for the MSSM.}
\label{tab:MSSM}
\end{table}

\subsection{NMSSM}
In the NMSSM, The particle content of the MSSM is extended by a gauge singlet superfield $\hat S$.  The most general, renormalisable and $R$-parity conserving superpotential reads
\begin{equation}
\label{eq:GNMSSM}
 {W}_S = {W}_{\rm{Yukawa}}  + \frac{1}{3}\kappa \hat S^3+ {\mu \hat H_u \hat H_d} + \lambda \hat S \hat H_u \hat H_d +  {\frac{1}{2} \mu_s \hat S^2 + t_s \hat S}  \;,
\end{equation}
The dimensionful terms can be forbidden by a $Z_3$ symmetry. In general, the scalar sector consists of {3 CP-even} and {2 CP-odd} states in the CP conserving limit, respectively 5 mixed states. An overview about the available (stand-alone) tools for the NMSSM is given in Tab.~\ref{tab:NMSSM}. One can see that so far no EFT techniques are used in the NMSSM codes, i.e. one should be careful and not use very large SUSY masses.

\begin{table}[tb]
\begin{center}
\begin{tabular}{|p{3.0cm}|  >{\centering\arraybackslash}p{3.0cm}  >{\centering\arraybackslash}p{3.0cm} >{\centering\arraybackslash}p{3.0cm} |}
\hline 
                    &  {\tt NMSSMCalc}\cite{Baglio:2013iia}   & {\tt NMSSMTools} \cite{Ellwanger:2004xm}     &  {\tt SoftSUSY}\cite{Allanach:2013kza}       \\
\hline  
\hline 
\multicolumn{4}{|c|}{General} \\
\hline
No $Z_3$            &             \no                &    \yes              &  \yes     \\
GUT model           &             \no                &    \yes              &  \yes      \\
CPV                 &  \yes & \yes & \no \\
scheme                                    & OS, \DR             & \DR                                  & \DR   \\
\hline         
\hline
\multicolumn{4}{|c|}{Heavy SUSY} \\
\hline
supported           & \no & \no & \no \\
EFT                 & - & - & - \\  
\hline
\hline 
\multicolumn{4}{|c|}{Other observables} \\
\hline 
Decays                  & \yes      & \yes                                 & \yes              \\
Flavour           & \no               & \yes         & \no  \\
g-2           & \no               & \yes         & \no  \\
Oblique P.           & \no               & \yes         & \no  \\
EDMs               & \yes                & \yes            & \no \\
\hline
\end{tabular}
\end{center}
\caption{Public tools for the NMSSM.}
\label{tab:NMSSM}
\end{table}

\section{Non-SUSY Tools}
As already mentioned, masses are used as input for non-SUSY tools. Thus, their tasks are a bit different compared to SUSY tools and usually consist of:
\begin{itemize}
 \item Calculate couplings 
 \item Calculate decays 
 \item Check theoretical constraints: (i) Perturbative unitarity, (ii) Vacuum stability
\end{itemize}
The reason for checking perturbative unitarity is that randomly chosen masses could correspond to arbitrary large Lagrangian parameters. Therefore, 
unitarity for 2 $\to$ 2 scattering processes is imposed to filter unphysical points. However, the results very often used in literature are based on crucial assumptions. 
I'll discuss this in mode detail in sec.~\ref{sec:unitarity}. \\
An overview about the existing tools for non-SUSY models is given in Tab.~\ref{tab:nonsusy}.

\begin{table}[tb]
\begin{center}
\begin{tabular}{|p{3.0cm}| >{\centering\arraybackslash}p{3.0cm} >{\centering\arraybackslash}p{3.0cm} >{\centering\arraybackslash}p{4.0cm}   |}
\hline 
                    &  {\tt GMcalc} \cite{Hartling:2014xma}              & {\tt 2HDMC}\cite{Eriksson:2009ws} & {\tt HDECAY} extensions \cite{Contino:2014aaa,Costa:2015llh,Fontes:2017zfn,Engeln:2018mbg} \\
\hline  
\hline 
\multicolumn{4}{|c|}{General} \\
\hline
Models               &  Georgi-Machacek          &   2HDMs & EFT ({\tt eHDECAY}), SM+Singlet ({\tt sHDECAY}), 2HDM ({\tt C2HDM\_HDECAY}), N2HDM ({\tt N2HDECAY}) \\
\hline         
\hline
\multicolumn{4}{|c|}{Higgs Results} \\
\hline
CPV    & \no  & \no  & (\yes) \\
Decays & \yes & \yes & \yes  \\
\hline
\hline 
\multicolumn{4}{|c|}{Theoretical checks} \\
\hline 
Unitarity  & \yes & \yes & 2HDM \\
Vacuum stability & \yes & \yes  & 2HDM \\
\hline
\hline 
\multicolumn{4}{|c|}{Other Observables} \\
\hline 
Flavour & \no & \no & \no \\
Oblique Parameters & \yes & \yes & 2HDM\\
$g-2$ & \yes & \yes & \no\\
EDMs  & \no  & \no  & \no \\
\hline
\end{tabular}
\end{center}
\caption{Public tools for non-SUSY models.}
\label{tab:nonsusy}
\end{table}

\section{Multipurpose Tools}
The idea of multipurpose tools it to generate {\bf automatically} a numerical tool for the study of a given model. For this purpose, one needs expressions for 
masses, vertices, RGEs and loop corrections. This is exactly the purpose of the package {\tt SARAH}\cite{Staub:2008uz,Staub:2009bi,Staub:2010jh,Staub:2012pb,Staub:2013tta}: {\tt SARAH} is a Mathematica package which 
was developed to get from a minimal input all important properties of SUSY and non-SUSY models:
\begin{itemize}
\item all {Vertices}, {Tadpole equations}, {Masses} and {Mass matrices}  
\item {Two-loop RGEs} including the {full CP and flavour} structure, effects of {gauge kinetic mixing},
and {$R_\xi$ dependence of VEVs}.  
\item {Loop corrections} to masses and precision observables
\end{itemize} 
The necessary input to define a model are the gauge symmetries, particle content, (super)potential, and how the symmetries are broken. {\tt SARAH} is already delivered with a large variety of SUSY and non-SUSY models.
%
Nowadays, there are two interfaces which make use of the derived information to obtain a spectrum generator:
\begin{itemize}
 \item The {\tt SARAH}/{\tt SPheno} interface 
 \item {\tt FlexibelSUSY} \cite{Athron:2014yba,Athron:2017fvs}
\end{itemize}

\begin{table}[tb]
\begin{center}
\begin{tabular}{|p{2cm}| >{\centering\arraybackslash}p{6cm} >{\centering\arraybackslash}p{6cm}  |}
\hline 
                    &  {\tt FlexibleSUSY}               & {\tt SPheno}   \\
\hline         
\hline
\multicolumn{3}{|c|}{\bf Higgs Results} \\
\hline
one-loop & full & full \\
two-loop & \parbox{6cm}{(N)MSSM hardcoded \\ general results in preparation} & fully model dependent calculation   \\
\hline
\multicolumn{3}{|c|}{\bf  Heavy BSM scales} \\
\hline
supported & \yes & \yes \\
scenarios & several MSSM limits hardcoded & \parbox{6cm}{general one-loop matching}   \\
\hline
\hline 
\multicolumn{3}{|c|}{\bf  Theoretical constraints} \\
\hline 
Unitarity  & \no & \yes \\
Vacuum Stability  & \no & via {\tt Vevacious} \\
\hline
\hline 
\multicolumn{3}{|c|}{\bf  Other Calculations} \\
\hline 
Decays & in progress & \yes \\
Flavor  & \no & \yes \\
$g-2$   & \yes & \yes \\
EDMs   & \yes & \yes \\
Oblique P.   & ($\delta m_W$) & \yes \\
\hline
\end{tabular}
\end{center}
\caption{}
\end{table}

\begin{figure}[tb]
 \begin{center}
  \includegraphics[width=0.5\linewidth]{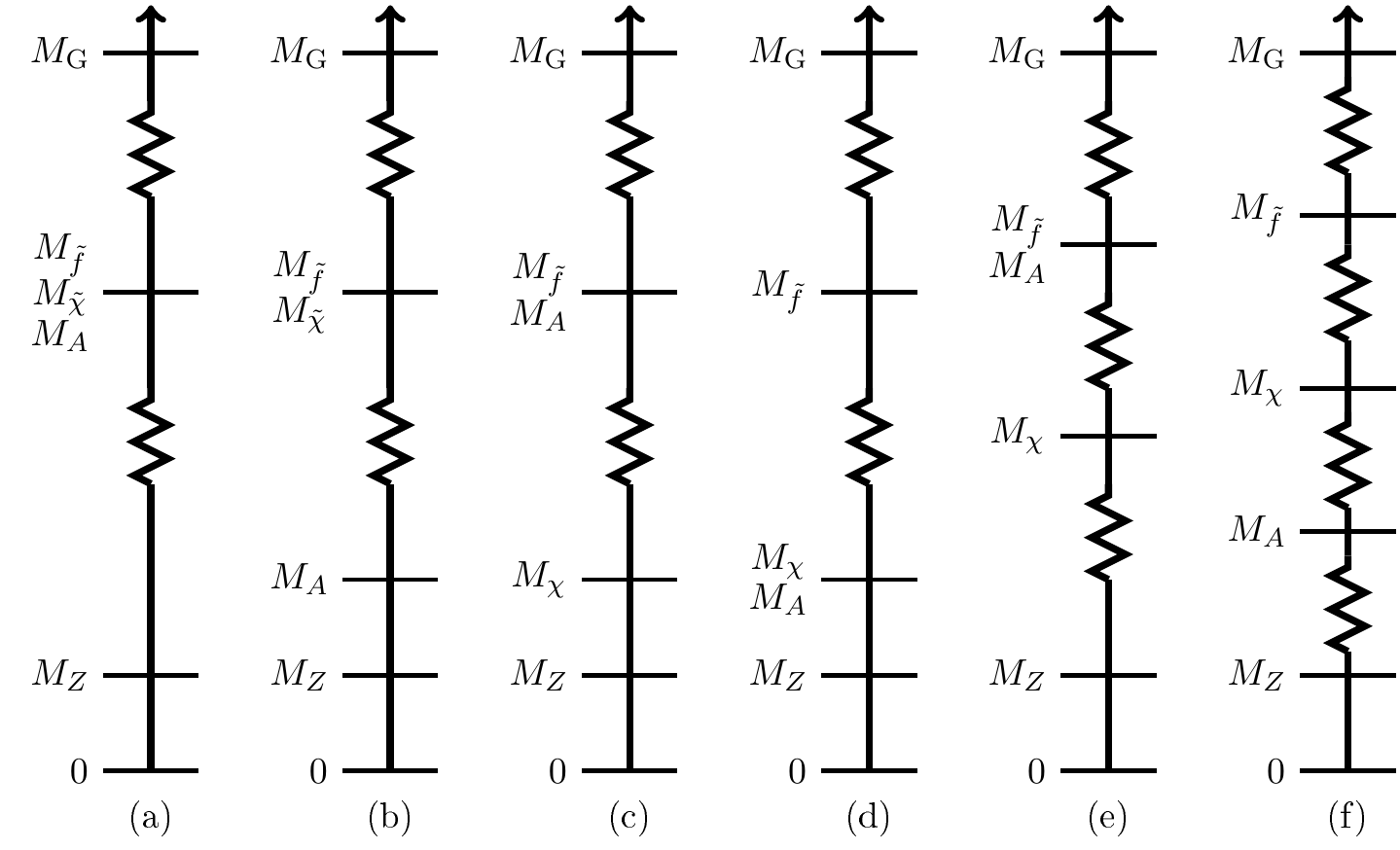}
 \end{center}
 \caption{Mass hierarchies for the MSSM now implemented in the \SARAH/\SPheno framework.}
 \label{fig:hierarchies}
\end{figure}

The obtained precision for scalar masses is comparable to the dedicated MSSM tools mentioned above. However, for other SUSY models the framework \SARAH/\SPheno is the only combination which provides important 
two-loop corrections. In the example for the NMSSM, these are the corrections $(\alpha_\lambda+\alpha_\kappa)^2$ for instance. Moreover, with \SARAH and \SPheno also loop corrected masses for non-SUSY models are 
available. This becomes important if UV completions of a model are considered: in that case, the input parameters are set at a scale well above the electroweak scale and the calculated masses at the weak scale are no longer an in- but an output, see e.g. \cite{Krauss:2018thf} for a discussion in the context of 2HDMs. \\
Until recently, \SARAH/\SPheno used a fixed order calculation which could be combined with a pole mass matching in the case that all particles but the SM ones are very heavy. This has been improved and an automatised matching between two scalar sectors is now available \cite{Gabelmann:2018axh}. This makes it possible to study all kind of mass hierarchies for a given UV model. The hierarchies already implemented in \SARAH for the example of the MSSM involving very heavy states are summarised in Fig.~\ref{fig:hierarchies}. With this setup, one could reproduce the results of {\tt SusyHD}, {\tt MhEFT} in the respective limits. 

\section{A few words about unitarity constraints}
\label{sec:unitarity}
\begin{figure}[tb]
 \begin{center}
  \includegraphics[width=0.5\linewidth]{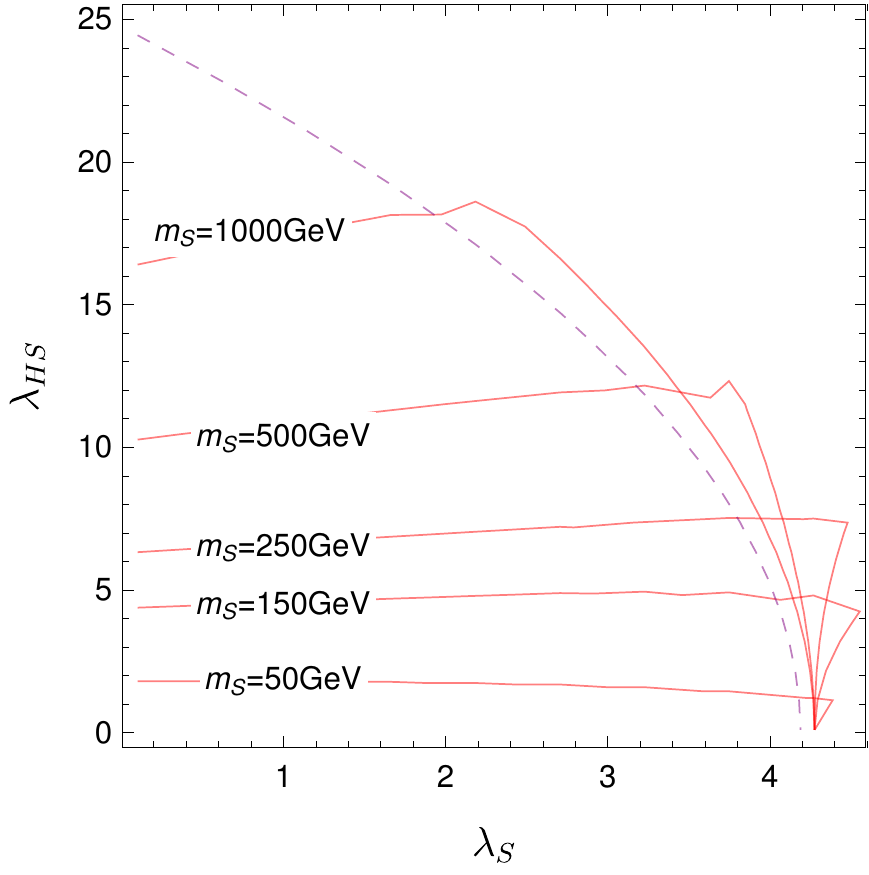}
 \end{center}
\caption{Constraints from perturbative unitarity on the quartic couplings in the singlet extended SM. The purple dashed lines shows the constraint when using the large $s$ approximation, while the red lines show the results from the full calculation. Plot taken from \cite{Goodsell:2018tti}.} 
\label{fig:SSM}
\end{figure}

We had mentioned that most non-SUSY tools check the validity of the underlying Lagrangian parameters. For doing that, they impose constraints on the the $2\to 2$ scattering cross sections:
the maximal eigenvalue $|(a^{\rm max}_0)|$ of the scattering matrix must fulfil 
\begin{equation}
(a^{\rm max}_0)| \leq \frac{1}{2} 
\end{equation}
Usually, the scattering matrices become huge, e.g. they have dimension $36\times 36$ in 2HDMs and $91\times 91$ in the Georgi-Machacek model. Therefore, very often the approximation is used that all involved masses are much smaller than the scattering energy, i.e. 
\begin{equation}
s \gg m_i^2 
\end{equation}
The consequence is that only point interactions contribute, while (effective) cubic couplings are completely neglected.  \\
In order to test the validity of this approximation, a full tree-level calculation is now available in \SARAH/\SPheno taking into also the propagator diagrams \cite{Goodsell:2018tti}. The difference compared the 
large $s$ approximation can be tremendous as shown in Fig.~\ref{fig:SSM} at the example of a singlet extensions. 

For scattering energies $s$ close to the masses, we find that the maximal eigenvalue of the scattering matrix is enhanced by a factor
\begin{equation}
\frac{|a^s_0|}{|a^{s\to\infty}_0|}\sim\frac{\lambda_{HS} v^2}{m_h^2} 
\end{equation}
Thus, the constrains on $\lambda_{HS}$ become stronger up to a factor of 10 for small singlet masses. \\
The setup was also used to check other models like 2HDMs \cite{Goodsell:2018fex} or triplet extensions \cite{Krauss:2018orw}. In all cases parameter regions were found, where the large $s$ approximations fails.

\section{Summary}
Precise  predictions for Higgs observables are crucial today. Therefore, many public tools were developed in the past for a fast and accurate study of Higgs properties in BSM models. While the stand-alone numerical tools concentrate on the most widely studied models (MSSM, NMSSM, 2HDMS, GM and SSM), so called 'Spectrum-Generators-Generators' can be used to go beyond that. These tools create automatically a spectrum generator for a given model which provides very similar features for a large variety of models as the well established tools for the most prominent ones.

\section*{Acknowledgements}
I thank the organisers of Charged2018 for the invitation and the hospitality during the workshop. 
I'm supported by the ERC Recognition Award ERC-RA-0008 of the Helmholtz Association.

\end{document}